\documentclass{ws-ijmpa}
\usepackage[super,compress]{cite}
\begin{document}

\markboth{P. K. Khandai, P. Sett, P.Shukla, V. Singh}
{ Hadron spectra in p+p collisions at RHIC and LHC energies}

\catchline{}{}{}{}{}
%
%%%%%%%%%%%%%%%%%%%%%%%%%%%%%%%%%%%%%%%%%%%%%%%%%%%%%%%%%%%%%%%%%%%%

\title{Hadron spectra in p+p collisions at RHIC and LHC energies}

\author{P. K. Khandai}
\address{Department of Physics, Banaras Hindu University, Varanasi, 221005, India 
pkkhandai@gmail.com}

\author{P. Sett}
\address{Nuclear Physics Division, Bhabha Atomic Research Center, Mumbai, 400085, India}

\author{P. Shukla}
\address{Nuclear Physics Division, Bhabha Atomic Research Center, Mumbai, 400085, India
pshukla@barc.gov.in}

\author{V. Singh}
\address{Department of Physics, Banaras Hindu University, Varanasi, 221005, India} 

\maketitle

\begin{history}
\received{Day Month Year}
\revised{Day Month Year}
\end{history}

\begin{abstract}
  We present the systematic analysis of transverse momentum ($p_T$) spectra of identified 
hadrons in p+p collisions at RHIC ($\sqrt{s}$ = 62.4 and 200 GeV) and at LHC energies 
($\sqrt{s}$ = 0.9, 2.76 and 7.0 TeV) using phenomenological fit functions.
  We review various forms of Hagedorn and Tsallis distributions and show their equivalence.
 We use Tsallis distribution which successfully describes the spectra in p+p collisions using 
two parameters, Tsallis temperature $T$ which governs the soft bulk spectra and 
power $n$ which determines the initial production in partonic collisions.
  We obtain these parameters for pions, kaons and protons 
as a function of center of mass energy ($\sqrt{s}$). It is found that the parameter $T$ has a 
weak but decreasing trend with increasing $\sqrt{s}$. The parameter $n$ decreases with 
increasing $\sqrt{s}$ which shows that production of hadrons at higher energies are 
increasingly dominated by point like qq scatterings. 
  Another important observation is with increasing $\sqrt{s}$, the separation 
between the powers for protons and pions narrows down hinting that the baryons and mesons are 
governed by same production process as one moves to the highest LHC energy.

\keywords{Tsallis distribution; Hadron spectra.}
\end{abstract} 

\ccode{PACS numbers: 13.85.Ni, 24.85.+p}

\section{Introduction}
  The heavy ion collisions at Relativistic Heavy Ion Collider (RHIC) and Large Hadron 
Collider (LHC) are performed to study strongly interacting matter at high energy 
density \cite{INTRO}.
 The measurements in $p+p$ collisions are important to understand particle production 
mechanism \cite{PPPROD} and are also used as baseline for heavy ion (HI) collisions.
  The hadron spectra provide insight into particle production as well as interaction 
in the hadronic and quark gluon plasma (QGP) phases.
  The hadron spectra at high transverse momentum ($p_T$) arise from 
fragmentation of high $p_T$ partons or jets from initial hard partonic collisions and
are known to follow a power law distribution. The high $p_T$ hadrons are very important for 
QGP studies as they measure jet quenching \cite{jet_quenching} effect in QGP. 
  The low $p_T$  hadrons form the bulk of the spectra arising from multiple scatterings 
and follow exponential distribution depictive of particle distribution in a thermal system 
which may be more meaningful for heavy ion system.  
 In addition, for heavy ion systems the hadron spectra at intermediate $p_T$ can 
arise from quark recombination \cite{Recom_Model1, Recom_Model2}.

  We review various forms of Hagedorn and Tsallis distributions which are 
used to describe p+p
collisions at RHIC and LHC energies and show their equivalence.
  The Tsallis distribution \cite{Tsallis,scale_ref} can reproduce the full spectral 
shape of hadrons with just two parameters; temperature $T$ and a non-extensive parameter 
$q$ which is related to a measure of temperature fluctuations and the degree of 
non-equilibrium in the system \cite{q_Tsallis}.
  The Tsallis functional form is mathematically same as Hagedorn 
distribution \cite{Hagedorn} with the parameter $q$ related to power $n$ of Hagedorn 
distribution which governs initial partonic collisions \cite{HAGEFACT}.
 The Hagedorn function has been successfully used to describe the meson spectra in p+p 
collisions at RHIC energies with an additional parameter to describe heavy ion collisions
as well \cite{mesonscaling}.

  In p+p collisions, the parameter $T$ governs soft collisions \cite{PPPROD} which is related to 
freeze out temperature for larger systems produced in relativistic heavy ion collisions
after collective effects are separated out. The parameter $n$ gives 
a good idea of initial production. 
 A value of $n$ close to 4 indicates point-like qq scattering (leading twist) while
a large value of $n$ is indicative of multiple scattering centers (higher twist effect) 
\cite{BlankenbeclerPRD12, BrodskyPLB637}. 
   A successful description of particle spectra with the Tsallis distribution allow us to 
calculate the integrated yield which provide important information on the bulk properties 
of the soft particle production and also used to infer the degree of chemical equilibration
when compared with the thermal model \cite{Thermal_model}.

  In the present work we use Tsallis distribution fit to draw systematics 
from the transverse momentum ($p_T$) spectra of identified hadrons measured
in p+p collisions at RHIC ($\sqrt{s}$ = 62.4 and 200 GeV) and at LHC energies 
($\sqrt{s}$ = 0.9, 2.76 and 7.0 TeV).  We obtain the Tsallis parameters 
$T$ and $n$ for pions, kaons and protons and study them as a function of 
center of mass energy.

\section{Particle spectra and the Tsallis distribution}
  The characteristics of transverse momentum spectra of particles produced in 
proton-proton or heavy ion collisions are keys to understand the particle 
production mechanisms.
 Hagedorn \cite {Hagedorn}, successfully described the shape of 
transverse mass $m_T$ spectra of hadrons produced in heavy ion collisions
in terms of two parameters by this form

\begin{eqnarray}
 E\frac{d^3N}{dp^3}  =  A \,\left(1 + \frac {m_{T}} {p_0}\right)^{-n}.
\label{Hageqn}
\end{eqnarray}
 Here $A$, $p_0$ and $n$ are fit parameters. The parameter $A$ is related to $dN/dy$ which we will show
little later.  
At low transverse momenta it assumes an 
exponential form and at large transverse momenta it becomes a power law which
mimics "QCD inspired" quark interchange model \cite {HAGEFACT} as follows:
 
\begin{eqnarray}
\left(1 + \frac {m_{T}} {p_0}\right)^{-n}
 & \simeq &  {\rm exp}\left(\frac {-nm_{T}} {p_0}\right), \, \, \, \,\,  {\rm for}\,\,\, p_{T} \rightarrow 0 \\
  & \simeq  &\left(\frac  {m_{T}} {p_0} \right)^{-n},  \,\,\,\,\,\,{\rm for}\,\,\, p_{T} \rightarrow \infty. 
\end{eqnarray}  
  The Hagedorn function has been successfully used to describe the meson spectra in p+p 
collisions at RHIC energies with an additional parameter to describe heavy ion collisions
as well \cite{mesonscaling}.

  The Tsallis distribution \cite{Tsallis,scale_ref} describes a thermal system 
in terms of two parameters $T$ and $q$ and is given by
\begin{equation}
G_q(E)  =  C_q \left(1+(q-1)\frac{E}{T}\right)^{-1/(q-1)}.
\label{BGeqn}
\end{equation}
Here $C_{q}$ is the normalization constant, $E$ is the particle energy,
$T$ is the Tsallis temperature and $q$ is the so-called nonextensivity parameter 
which measures the temperature fluctuations \cite{q_Tsallis} in the system 
as: $q-1 = Var(T)/<T>^{2}$.
 The values of $q$  lie between $1 < q < 4/3$. For $q$ $\rightarrow$ 1, 
the distribution corresponds to an equilibrated system described by a pure exponential 
(Boltzmann-Gibbs) type distribution \cite {BG_statistics1, BG_statistics2}:

\begin{equation}
G_q(E) \rightarrow  C_1 \, {\rm exp} \left(-\frac{E}{T}\right).
\end{equation}  
 Using the relations $E = m_{T} = \sqrt{p_{T}^{2} + m^{2}}$ at mid-rapidity
and $ 1/(q-1) = n$, Eq.~\ref{BGeqn} takes the form
\begin{eqnarray}
 E\frac{d^3N}{dp^3}  =  C_{n}\left(1 + \frac {m_{T}}{nT}\right)^{-n},
\label{Hageqn1}
\end{eqnarray}
which is same as Eq.~\ref{Hageqn} with $p_0 = nT$.
  Larger values of $q$ correspond to smaller values of $n$ describe a 
system away from thermal equilibrium. In terms of QCD, smaller values of
$n$ imply dominant hard point-like scattering. Phenomenological studies suggest that, 
for  quark-quark point scattering, $n\sim$ 4 \cite{BlankenbeclerPRD12, BrodskyPLB637}, 
and when multiple scattering centers are involved $n$ grows larger and can go upto 
20 for protons. 

   In order to associate the Tsallis distribution with a probability 
distribution, which describes the invariant particle spectra defined 
over $0 < E < \infty$, Eq.~\ref{Hageqn1} must satisfy a normalization and 
energy conservation condition. Using the unit normalization 
condition, we can determine the co-efficient $C_{n}$,
 the resulting formula used for fitting the hadron spectra used by
PHENIX \cite{PPG099} and is given by

\begin{eqnarray}\label{fitfun}
  E\frac{d^3N}{dp^3}  & = & \frac{1}{2\pi} \frac{dN}{dy} \frac{(n-1)(n-2)}{(nT+m(n-1))(nT+m)} \left(\frac{nT+m_T}{nT+m}\right)^{-n} % \nonumber\\
\end{eqnarray}
  In some of the papers \cite{scale_ref, Cleymans}, another variant of 
Tsallis distribution is written as

\begin{equation}
G_q(E)  =  C_q \left(1+(q-1)\frac{E}{T}\right)^{-q/(q-1)}.
\label{BGeqn2}
\end{equation}
 The corresponding normalized distribution in terms of $n$ can be obtained by 
replacing $n$ by $(n-1)$ in Eq.~\ref{fitfun}.

  The Tsallis distribution used by some publications of 
PHENIX \cite {pp62chargedpionkaonproton},
STAR \cite{ppproton, ppbaryon} and CMS \cite{IdentifiedHadronsCMS}
is of the form given by 

\begin{eqnarray}\label{fitfun2}
  E\frac{d^3N}{dp^3}  & = & \frac{1}{2\pi} \frac{dN}{dy} \frac{(n-1)(n-2)}{(nC+m(n-2))(nC)} \left(1 + \frac{m_T - m}{nC}\right)^{-n} % \nonumber\\
\end{eqnarray}
 This is same as Eq.~\ref{fitfun}, with the parameter $C$ of Eq.~\ref{fitfun2} 
related to $T$ by $nC$   $\longrightarrow$  $ nT + m$. In summary, all the forms
of Hagdorn and Tsallis distribution described above are essentially same with 
parameter of one form is related to the parameter of the other form by a definite 
relation. We will use the form given in Eq.~\ref{fitfun} in rest of the analysis.

\section{Results and Discussions}

In the present work, the measured $p_T$ distribution of hadrons are taken from 
PHENIX ($|y| < 0.35 $) \cite{PPG099, pp62chargedpionkaonproton, pp62pion}, 
STAR  ($|y| < 0.5 $, $|y| < 0.75 $) \cite{ppproton, ppbaryon},   
and CMS ($|y| < 1.0$) \cite{IdentifiedHadronsCMS} experiments.
 The average yields of charged pions, charged kaons and protons
measured at all RHIC ($\sqrt{s}$ = 62.4 and 200 GeV) and LHC energies 
($\sqrt{s}$ = 0.9, 2.76 and 7.0 TeV) are used in the analysis.
 The errors on the data are taken as quadratic sums of statistical 
and uncorrelated systematic errors wherever available.

\begin{figure}[pb]
\centerline{\psfig{file=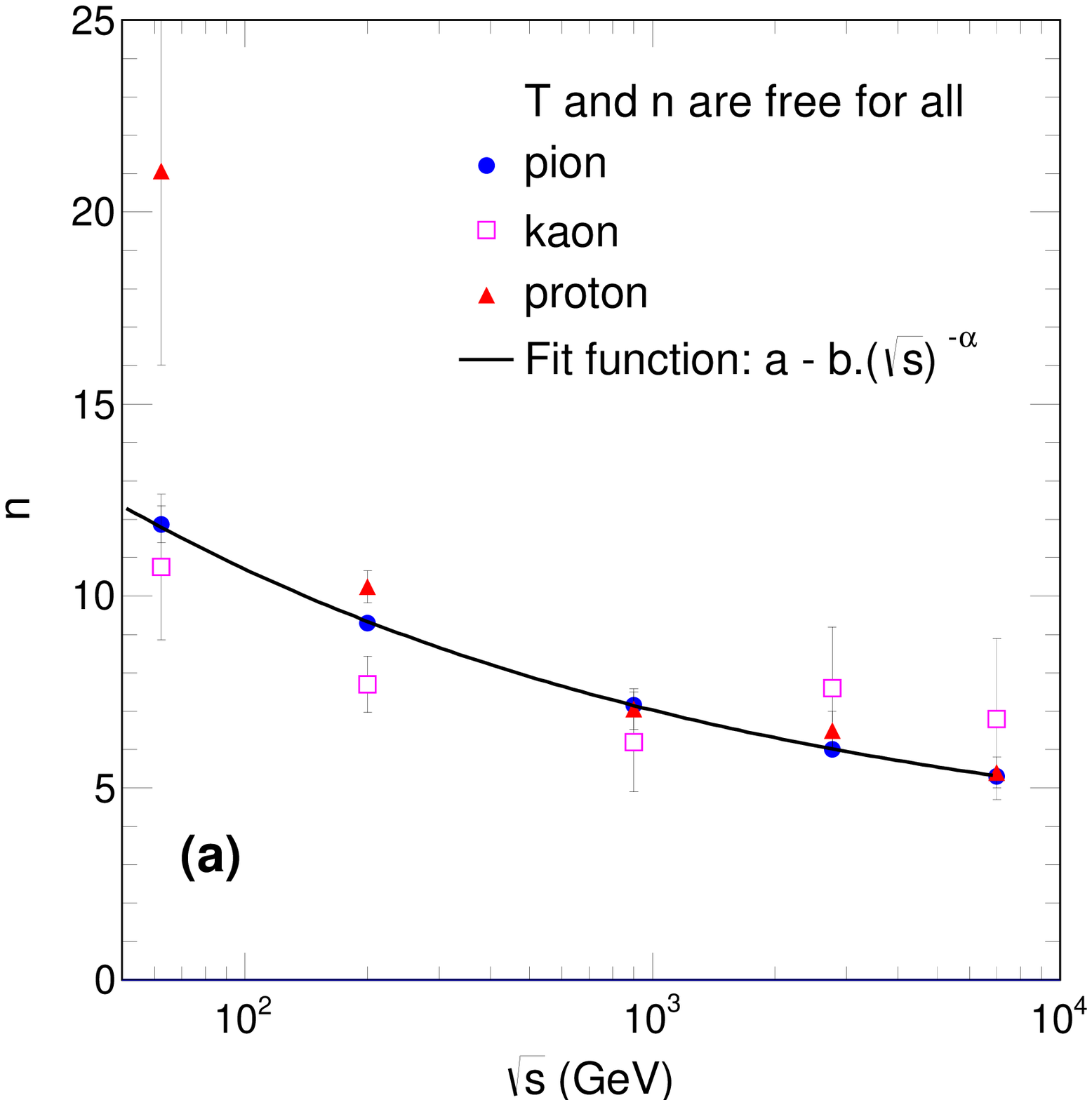,width=7.5cm}}
\centerline{\psfig{file=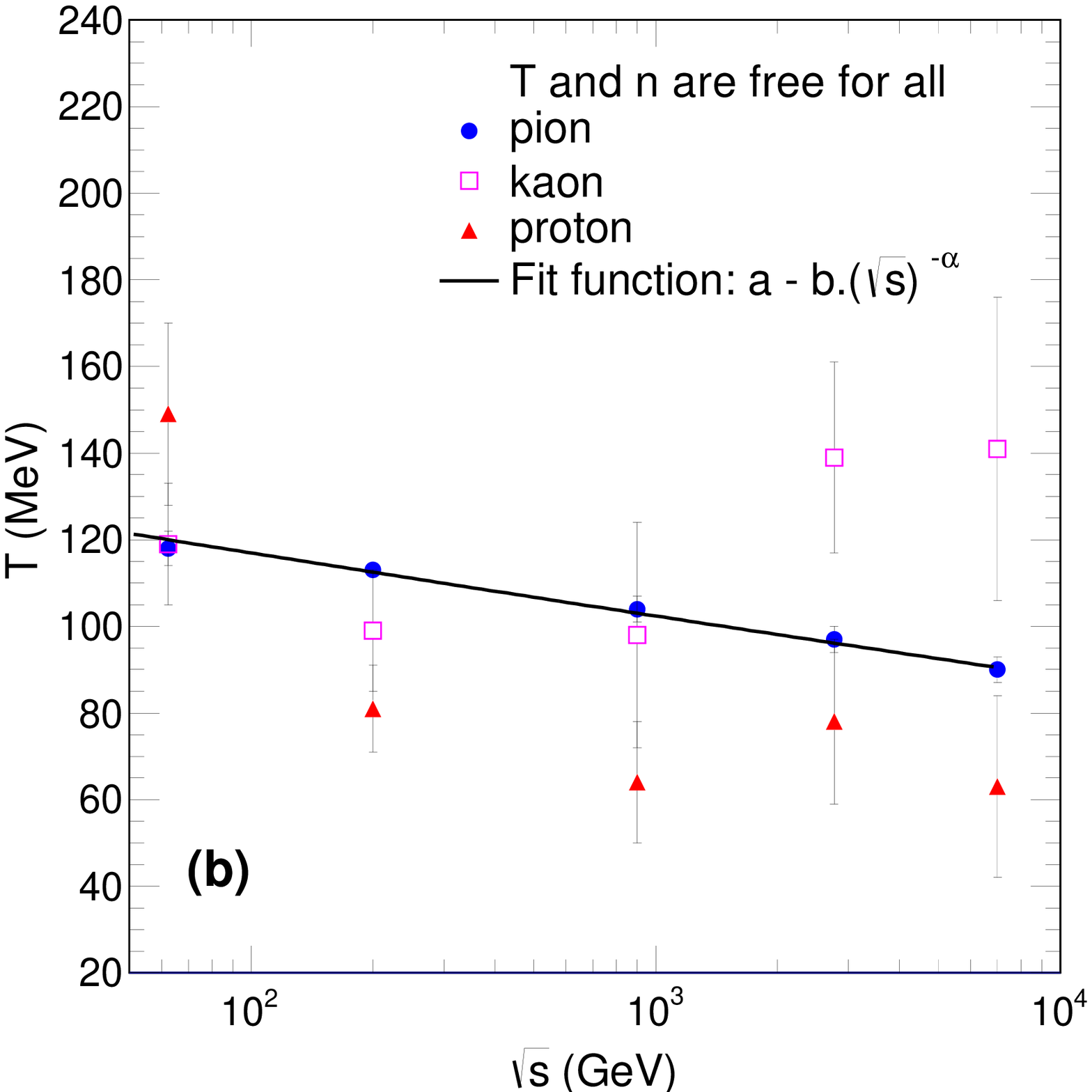,width=7.5cm}}
\vspace*{8pt}
\caption{The variation of Tsallis parameters (a) $n$ and (b) $T$ as a function 
of center of mass energy ($\sqrt{s}$) for pion, kaon and proton.
The solid curve in each figure are the function 
$a - b\,(\sqrt{s})^{-\alpha}$.\label{tsallis_fit_T_n_free}} fitted over all particles.
\end{figure}

  First, we fit all the measured hadron (pion, kaon and proton) spectra with 
Tsallis distribution (Eq.~\ref{fitfun}) keeping both the parameters $T$ and $n$ 
as free. Figure~\ref{tsallis_fit_T_n_free} shows the variation of parameters $T$ and $n$ 
as a function of $\sqrt{s}$ along with their parameterizations by a 
function: $a - b(\sqrt{s})^{-\alpha}$.
%The values of $T$ and $n$ for pions, kaons and protons at different $\sqrt{s}$ 
%are given in Table \ref{table1}.
  We notice that the parameter $T$ has a decreasing but weak dependence on $\sqrt{s}$
and the parameter $n$ has a definite decreasing trend with increasing $\sqrt{s}$.
 In case of pions, both the parameters vary smoothly as a function of collision energy.
  The solid line in Fig.~\ref{tsallis_fit_T_n_free}(a) is obtained by 
fitting $n$ for all particles with a single parameterized form. 
The solid line in Fig. ~\ref{tsallis_fit_T_n_free} (b) is the same for $T$
for all particles. These curve mostly pass through the pion points. 

  There is correlation between parameters $T$ and $n$, if any of the two 
increases, the other also increases and thus we need to fix one of them by some 
method and study the behaviour of the other as a function of particle type 
and $\sqrt{s}$.
  The parameter $T$ as defined in Eq.~\ref{fitfun} can be assumed to be same 
for all particles at a particular energy. We can fix it to the value averaged over all 
particle types or simply use pion $T$ for all the particles. 
  Thus first we fit the measured pion spectra with Tsallis distribution
by keeping parameters $T$ and $n$ as free and then fix $T$ for kaon and 
proton spectra to get $n$.

\begin{figure}[pb]
\centerline{\psfig{file=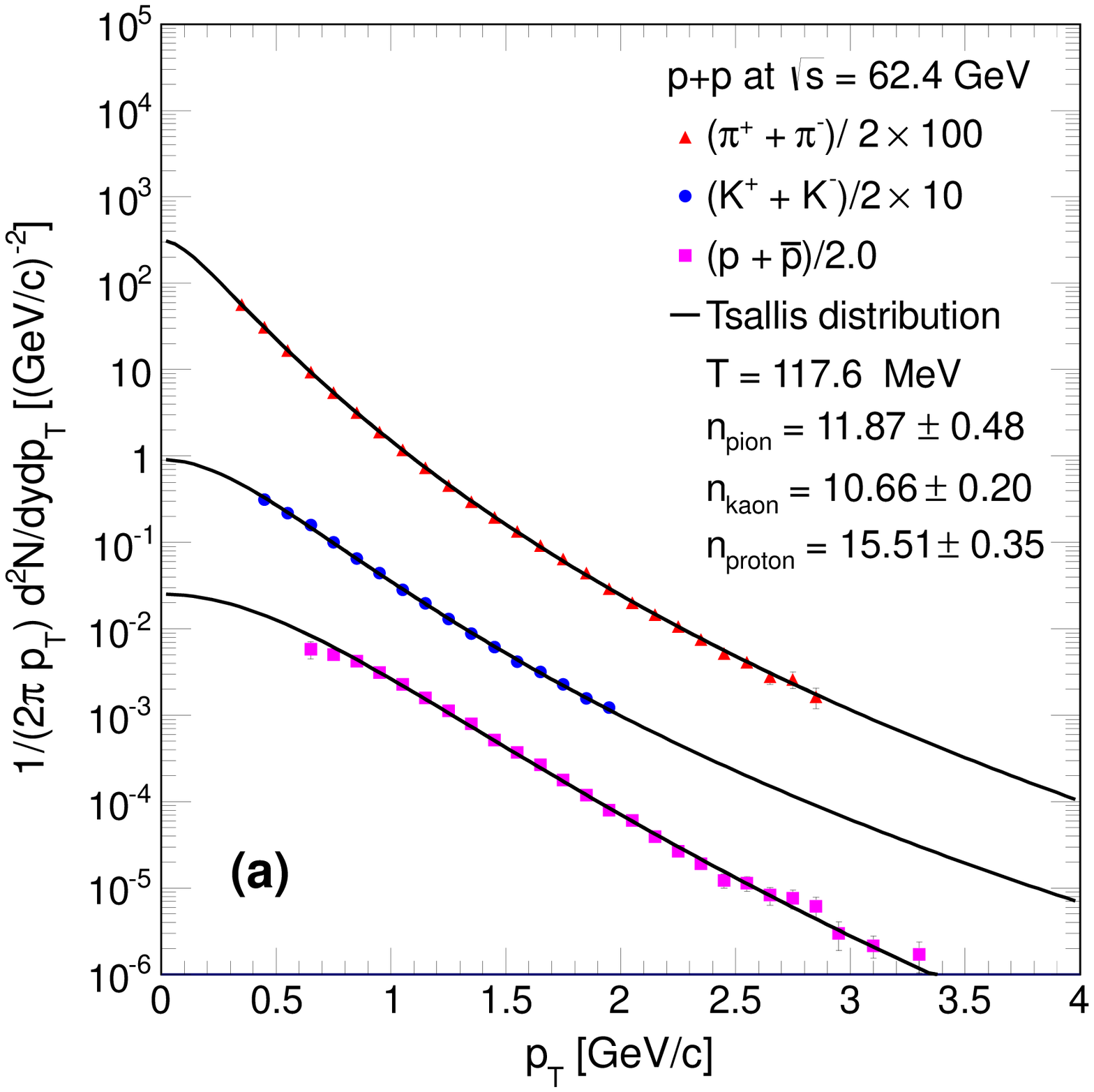,width=7.5cm}}
\centerline{\psfig{file=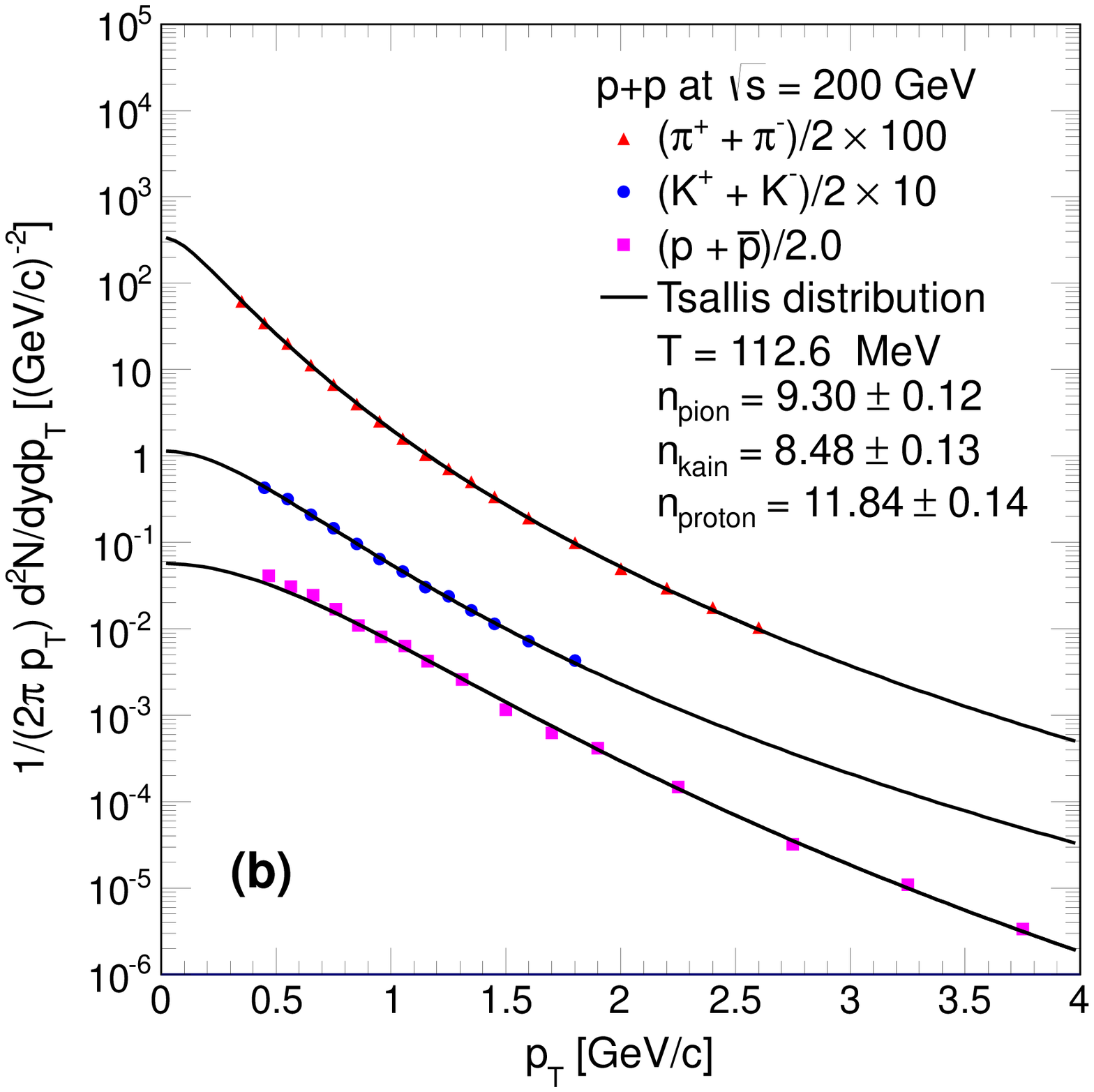,width=7.5cm}}
%\centerline{\psfig{file=allbaryons_pp.eps,width=7.5cm}}
\vspace*{8pt}
\caption{The invariant yields of pions, kaons and protons as a function of $p_T$ at 
(a) $\sqrt{s}$ = 62.4 GeV \cite{pp62chargedpionkaonproton, pp62pion}, and (b) 
 $\sqrt{s}$ = 200 GeV \cite{PPG099, ppproton, ppbaryon} for p+p collisions.
The solid curves are the fitted Tsallis distributions. 
\label{ppmesonsbaryons@RHIC}}
\end{figure}

\begin{figure}
\centerline{\psfig{file=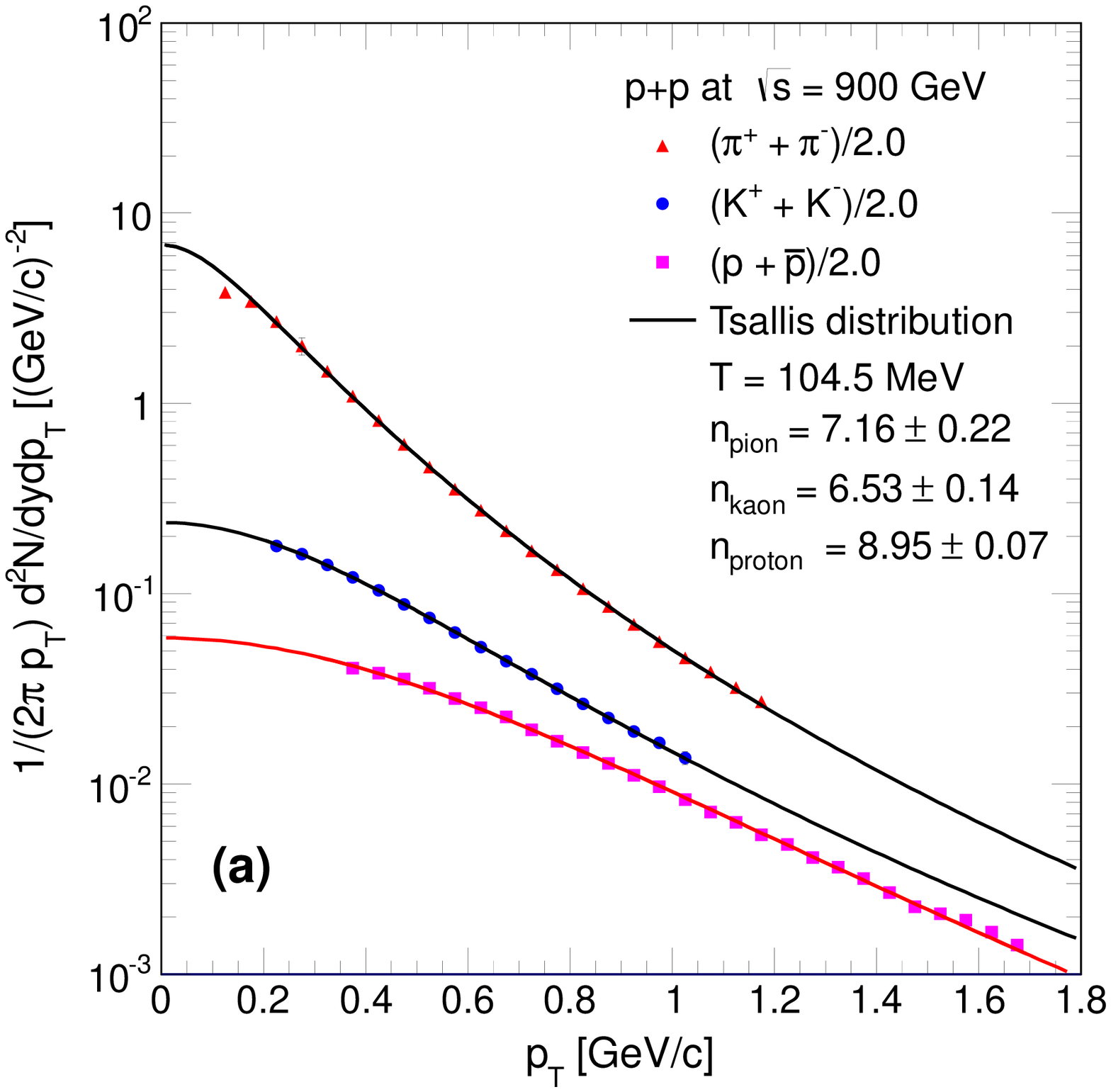,width=7.5cm}}
\centerline{\psfig{file=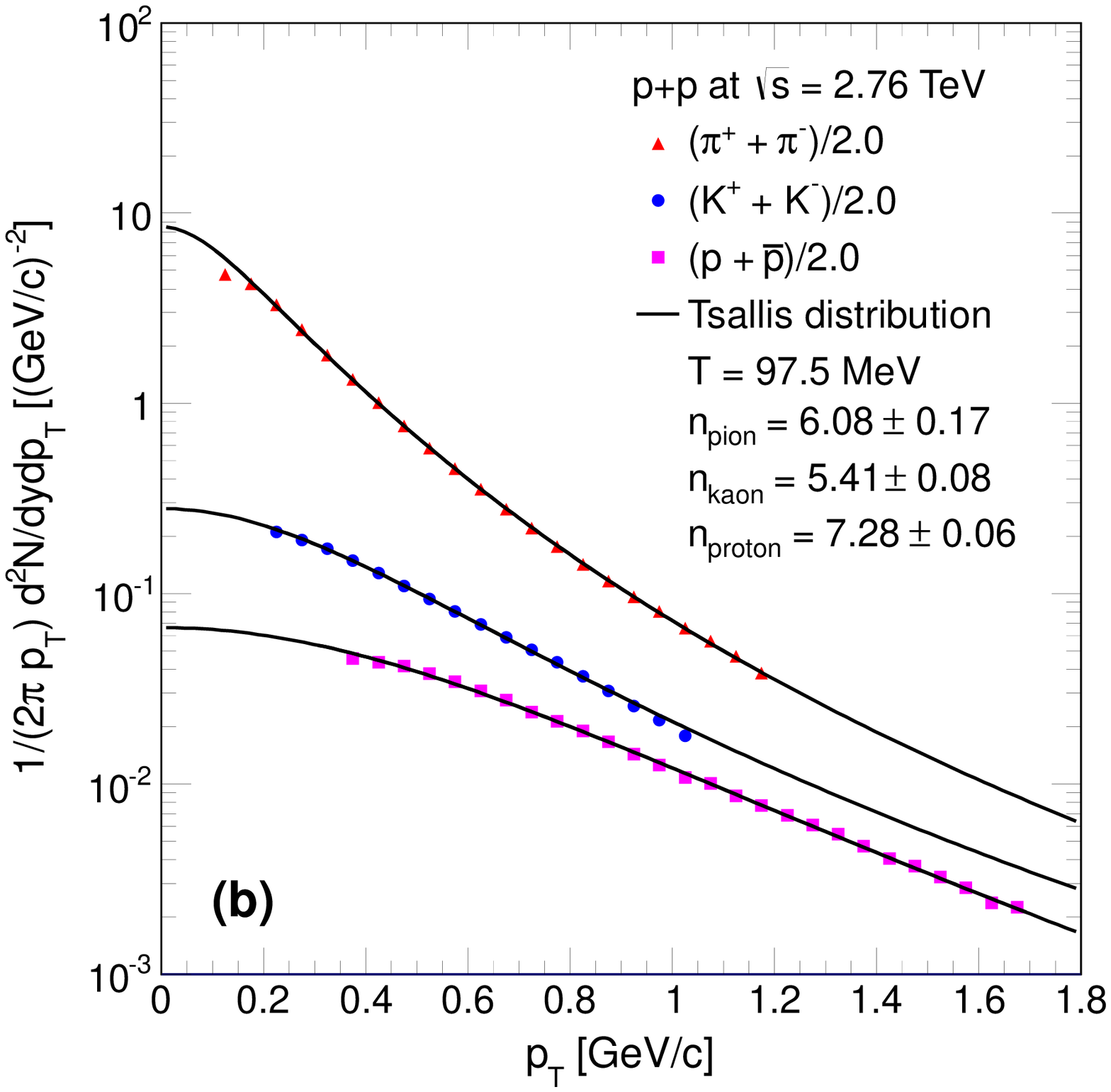,width=7.5cm}}
\vspace*{8pt}
\caption{The invariant yields of pions, kaons and protons as a function of 
as a function of $p_{T}$ for p+p collision at 
(a) $\sqrt{s}$ = 0.9 TeV \cite{IdentifiedHadronsCMS}, 
(b) $\sqrt{s}$ = 2.76 TeV \cite{IdentifiedHadronsCMS} and 
The solid curves are the fitted Tsallis distribution.}
\label{ppmesonsbaryons@LHC1}
\end{figure}

  Figure~\ref{ppmesonsbaryons@RHIC} shows the invariant yields of pions, kaons and protons 
as a function of $p_T$ for p+p collisions at
(a) $\sqrt{s}$ = 62.4 GeV \cite{pp62chargedpionkaonproton, pp62pion}, and 
(b) $\sqrt{s}$ = 200 GeV \cite{PPG099, ppproton, ppbaryon}.
   Figure~\ref{ppmesonsbaryons@LHC1} shows the invariant yields of pions, kaons and protons 
as a function of $p_T$ for p+p collisions at 
(a) $\sqrt{s}$ = 0.9 TeV \cite{IdentifiedHadronsCMS}, 
(b) $\sqrt{s}$ = 2.76 TeV \cite{IdentifiedHadronsCMS}.

\begin{figure}
\centerline{\psfig{file=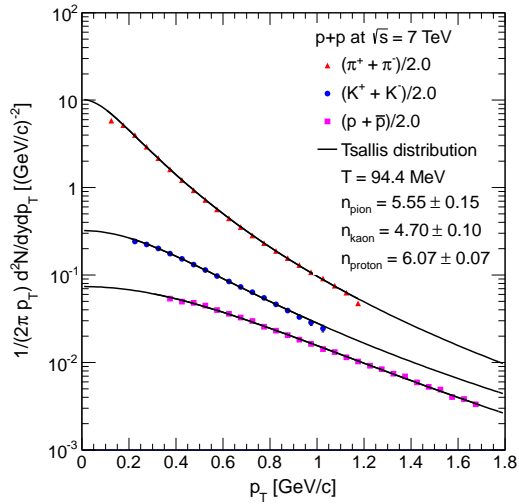,width=7.5cm}}
\vspace*{8pt}
\caption{The invariant yields of pions, kaons and protons as a function of 
as a function of $p_{T}$ for p+p collision at 7.0 TeV \cite{IdentifiedHadronsCMS}. 
The solid curves are the fitted Tsallis distribution.}
\label{ppmesonsbaryons@LHC2}
\end{figure}

  Figure~\ref{ppmesonsbaryons@LHC2} shows the invariant yields of pions, kaons and protons 
as a function of $p_T$ for p+p collisions at 7.0 TeV \cite{IdentifiedHadronsCMS}.

\begin{figure}
\centerline{\psfig{file=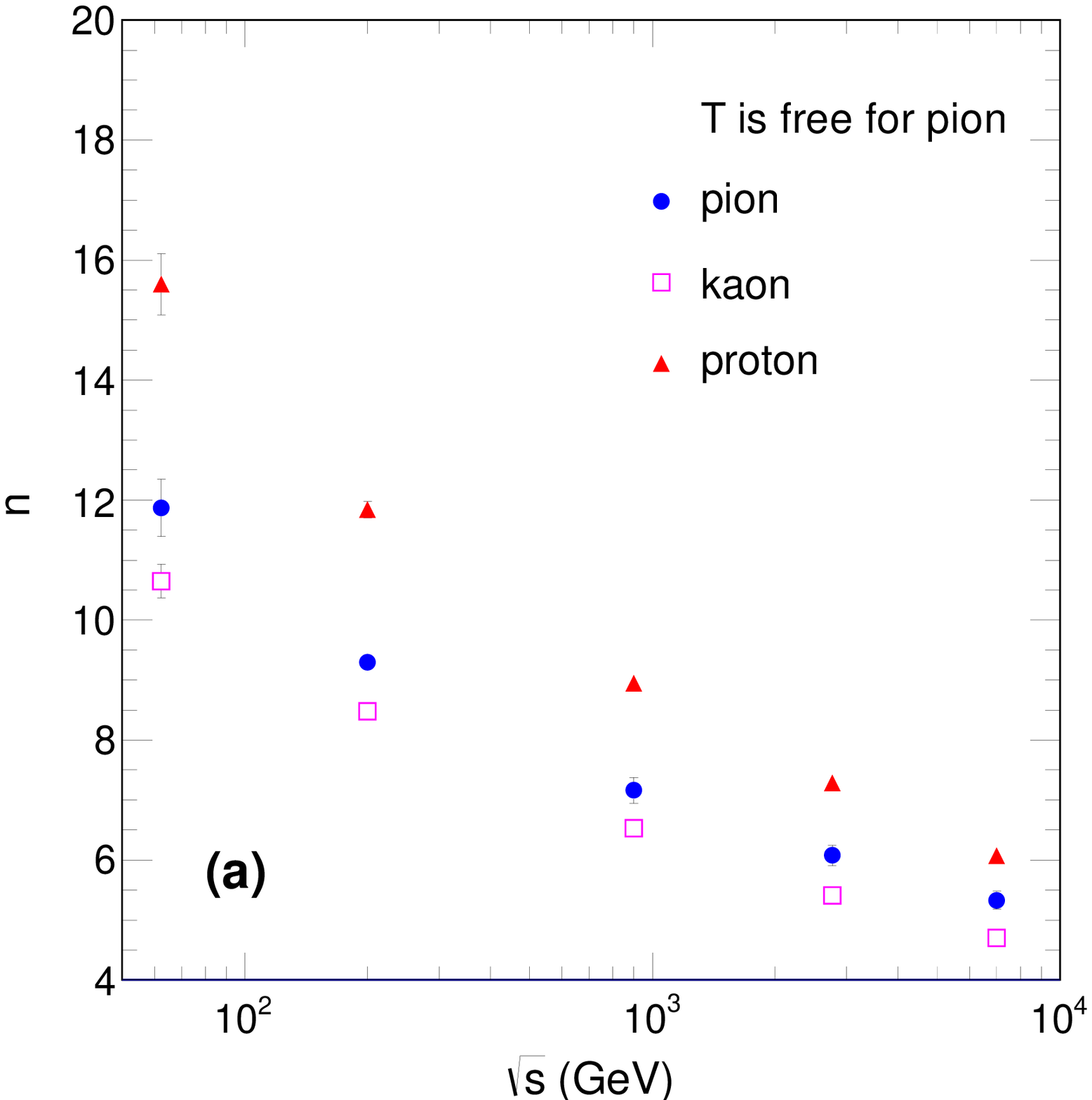,width=7.5cm}}
\centerline{\psfig{file=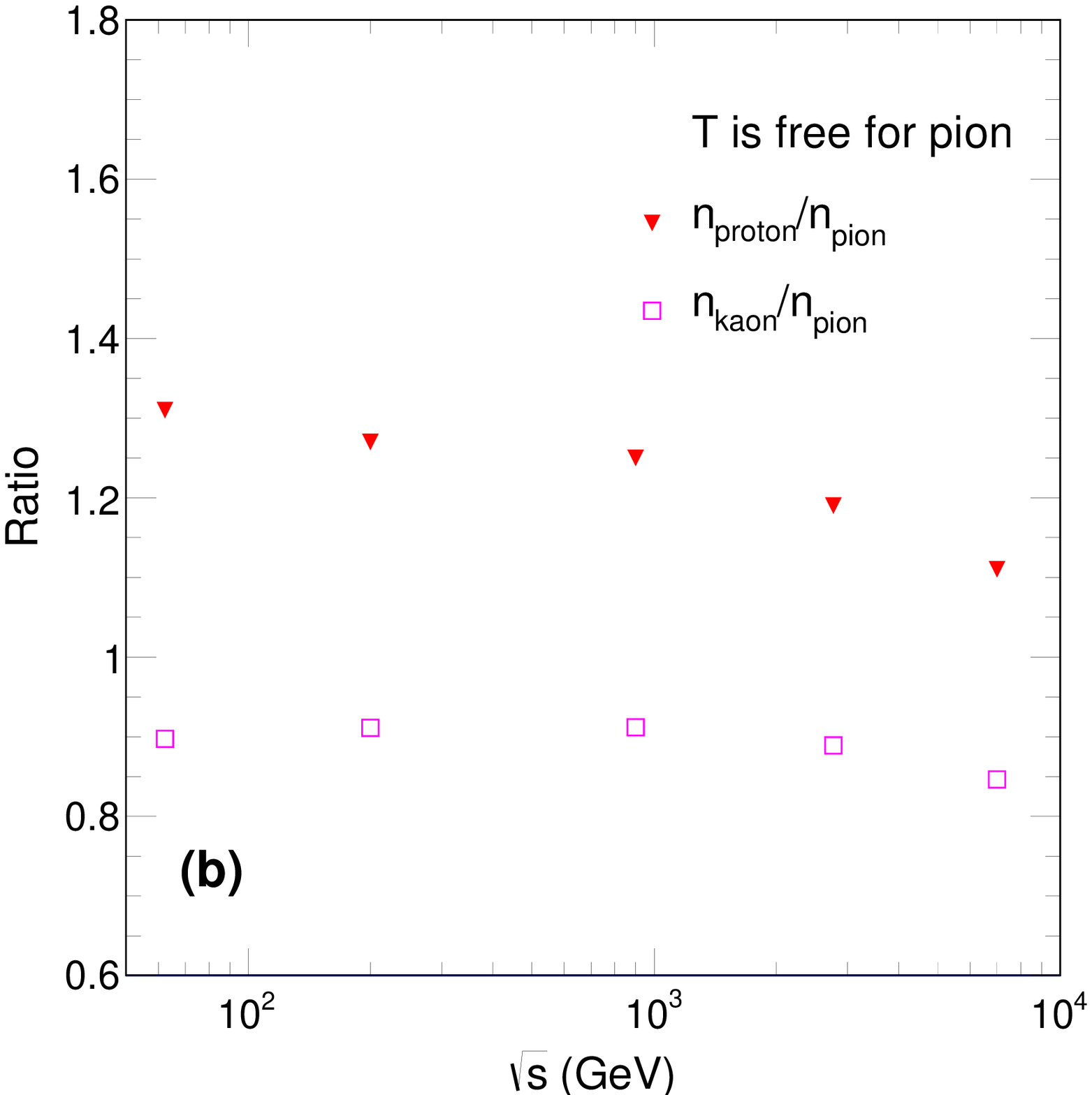,width=7.5cm}}
\vspace*{8pt}
\caption{(a) shows the variation of Tsallis parameter $n$ as a function of $\sqrt{s}$.
Here the value of parameter $T$ obtained from fitting pion spectra has been used in kaon 
and proton spectra. 
 and (b) shows the variation of $n_{proton}/n_{pion}$ and $n_{kaon}/n_{pion}$ as a function of 
$\sqrt{s}$. 
\label{tsallis_fit_n_free}}
\end{figure}

  The solid curves in all the figures are the fitted Tsallis Distribution of Eq.~(\ref{fitfun}).
  In case of pions both $T$ and $n$ are kept free during fits. 
For kaons and protons the values of $T$ is fixed to the value obtained for pions.
  The values of $T$ and $n$ for pions and $n$ for kaons and protons so obtained 
are given in Table \ref{table2} at different collision energies.
  It can be noted that that we get good $\chi^{2}$ values for fits at all energies 
with such procedure.
 In Fig~\ref{tsallis_fit_n_free}, upper plot (a) shows the variation of Tsallis parameter $n$ 
and the lower plot (b) shows the variation of ratio $n_{proton}/n_{pion}$ and $n_{kaon}/n_{pion}$ 
as functions of $\sqrt{s}$.
  It is observed that the parameter $n$, for pions, kaons and protons, 
is monotonically decreasing with 
increasing $\sqrt{s}$, which could be understood in terms of changing production
mechanism at different collision energies.

%%%%%%%%%%%%%%%%%%%%RHIC energies (yield vs pT)%%%%%%%%%%%%%%%%%%%

%%%%%%%%%%%%%%%%%%%%LHC energies (yield vs pT)%%%%%%%%%%%%%%%%%%%

%%%%%%%%%%%%%%%%%%%%%%%%%%%%%%%%%%%%%%%%%%%%%%%%%%%%%%%%%%

 The QCD cross sections scale as $1/p_T^n$ in terms of the power $n=2\,n_a - 4$ where
$n_a$ is the number of participating quarks \cite{BlankenbeclerPRD12, BrodskyPLB637}.
 If the dominant process in meson or baryon production is point like scattering 
$q q \rightarrow q q$ (referred as leading twist) the number of 
participating quarks is 4 and hence $n=4$. The power can go up due to 
contribution of higher twists processes explained in the following.
  The mesons scattering on quark $q \pi \rightarrow q \pi$ gives $n=2 \times 6 - 4 = 8$. 
For proton production if the subprocess $qq \rightarrow qqq\bar{q}$ is dominant then 
$n=8$. Proton scattering process on quark ($qp \rightarrow qp$) gives $n=12$.
 The value of $n$ for pions could go up to 16 for process $p\pi \rightarrow p\pi$
and that for protons could be 20 corresponding for process $pp \rightarrow pp$.
 
 The decreasing value of parameter $n$, for pions, kaons and protons 
with collision energy in Table \ref{table2} indicates that the dominant 
production process is moving from higher twist to leading twist as one goes up in $\sqrt{s}$. 
 We find that the separation of $n$ among pions, kaons and protons decreases as we go from 
lower center of  mass energy (at RHIC) to higher center of mass energy (at LHC). 
This indicates that at higher collision energies the production process is more like 
$qq$ point scattering for all particles which gives lower and similar values of $n$
for different particles.

\begin{table}[ph] 
\tbl{The values of Tsallis Parameters ($T$ and $n$) for pions, kaons and protons.
In case of pions both $T$ and $n$ are kept free during fits. For kaons and protons 
the values of $T$ is fixed to the value obtained for pions.} 
{\begin{tabular}{@{}ccccc@{}} \toprule 
\hline
$\sqrt{s}$ (TeV)  & $T$ (MeV)  &  $n$  &  $\frac{dN}{dy}$  &  $\chi^{2}/ndf$  \\
\hline
\multicolumn{5}{c}{For pion}                  \\
\hline
0.062   &  117.6 $\pm$ 1.2   & 11.87 $\pm$ 0.48  & 0.812 $\pm$ 0.03   &  0.19    \\
0.2     &  112.6 $\pm$ 2.1   & 9.30 $\pm$ 0.12   & 9.204 $\pm$ 0.05   &  4.52    \\
0.9     &  104.5 $\pm$ 3.1   & 7.16 $\pm$ 0.22   & 1.917 $\pm$ 0.014  &  1.18    \\
2.76    &   97.5 $\pm$ 2.6   & 6.08 $\pm$ 0.17   & 2.44 $\pm$ 0.02    &  1.04    \\
7.0     &   94.4 $\pm$ 2.8   & 5.55 $\pm$ 0.15   & 3.083 $\pm$ 0.03   &  0.88    \\
\hline
\multicolumn{5}{c} {For Kaon}                                         \\
\hline
0.062   &  117.6    & 10.66 $\pm$ 0.28  & 0.072 $\pm$ 0.002  &   0.17    \\
0.2     &  112.6    & 8.48 $\pm$ 0.13   & 0.493 $\pm$ 0.008  &   0.68    \\
0.9     &  104.5    & 6.53 $\pm$ 1.40   &  0.238 $\pm$ 0.002 &   0.07    \\
2.76    &  97.5     & 5.41 $\pm$ 0.08   &  0.322 $\pm$ 0.000 &   0.74    \\
7.0     &  94.4     & 4.70 $\pm$ 0.10   &  0.418 $\pm$ 0.006 &   0.33    \\
\hline
\multicolumn{5}{c} {For proton}                                     \\
\hline 
0.062   &  117.6    & 15.51 $\pm$ 0.35  & 0.007 $\pm$ 0.00   &   0.35    \\
0.2     &  112.6    & 11.84 $\pm$ 0.14  & 0.083 $\pm$ 0.003  &   1.18    \\
0.9     &  104.5    & 8.95 $\pm$ 0.07   & 0.105 $\pm$ 0.00   &   1.06    \\
2.76    &   97.5    & 7.28 $\pm$ 0.06   & 0.137 $\pm$ 0.001  &   1.35    \\
7.0     &   94.4    & 6.07 $\pm$ 0.07   & 0.177 $\pm$ 0.001  &   1.05    \\      
\hline 
\end{tabular}  \label{table2} }
\end{table}

%%%%%%%%%%%%%%%%%%%%%%%%  (THIRD RESULTS) %%%%%%%%%%%%%%%%%%%%%%%

\begin{figure}
\centerline{\psfig{file=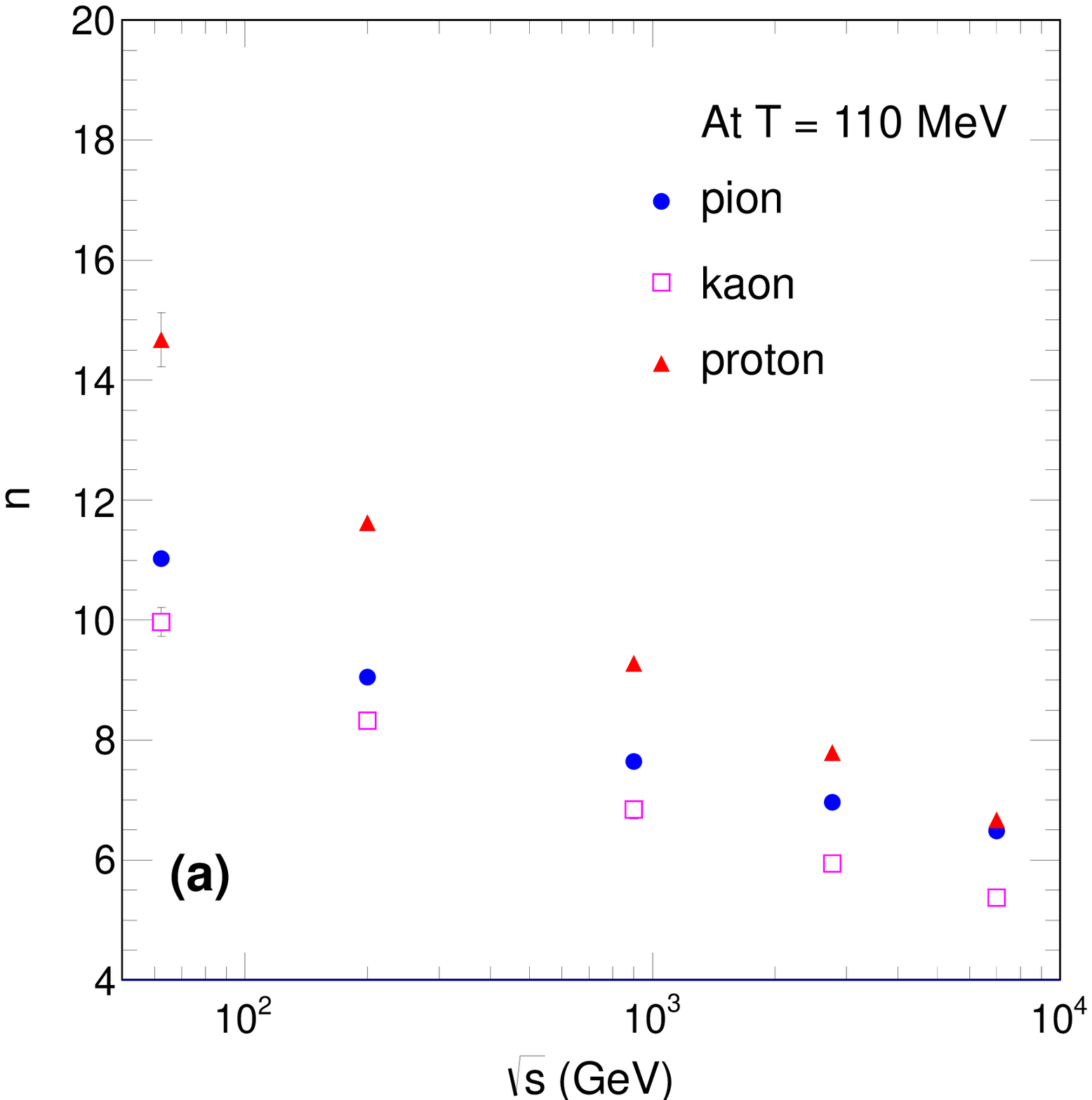,width=7.5cm}}
\centerline{\psfig{file=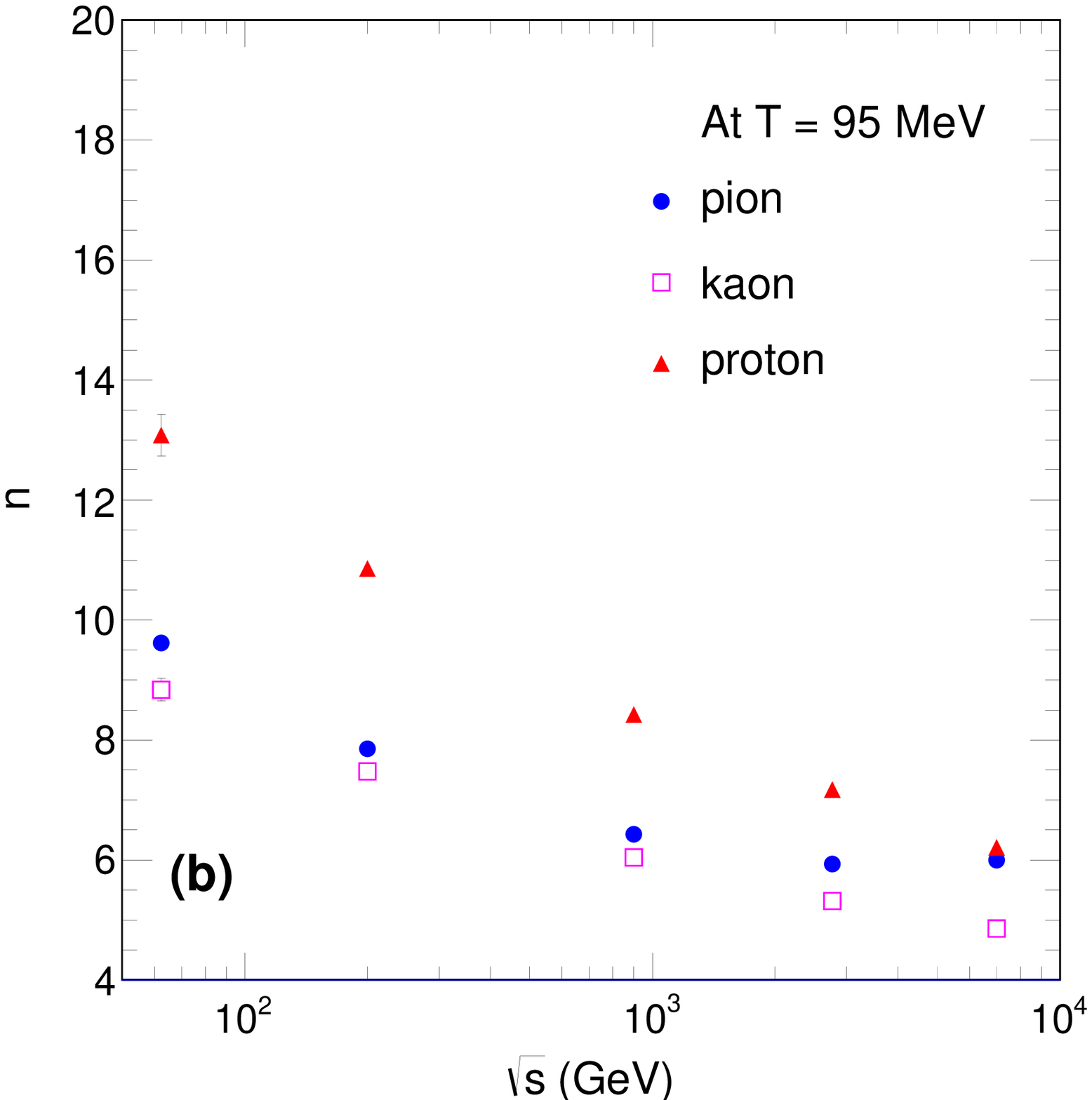,width=7.5cm}}
\vspace*{8pt}
\caption{The variation of Tsallis parameter $n$ as a function of $\sqrt{s}$, 
when $T$ is kept fixed at (a) 110 MeV and 
(b) at 95 MeV.\label{tsallis_fit_diff_T_fixed}}
\end{figure}

  Lastly, we study all hadron spectra at all energies using fixed values of $T$.
From the free fit results given in Fig~\ref{tsallis_fit_T_n_free}(b)
we choose two values  $T$ = 110 MeV and $T$=95 MeV and fit all hadron spectra to obtain 
power $n$ for pions, kaons and protons as a function of $\sqrt{s}$. The results for
$T$ = 110 MeV and 95 MeV are given in Figs.~\ref{tsallis_fit_diff_T_fixed}(a) 
and (b) respectively.
  We find that for both the values of $T$, the parameter $n$ as a function of 
$\sqrt{s}$ decreases which is noticed in all our analysis.
  Although the values of $n$ at $T$ = 110 MeV are greater than their respective 
values at $T$ = 95 MeV the general conclusions about $n$ remain the same.

\section{Conclusion}

  In the present work we use Tsallis fit to draw systematic trends 
from the transverse momentum spectra of identified hadrons measured
in p+p collisions at RHIC and LHC energies.
 We also review various forms of Hagedorn and Tsallis distributions and show their 
equivalence.
 We obtain the Tsallis parameters $T$ and $n$ for pions, kaons and protons and 
study them as a function of center of mass energy.
 Since $T$ and $n$ are correlated we fix the value of $T$ for all particles 
at a particular energy to the pion temperature.  
  In general, the Tsallis temperature $T$ has a decreasing but weak dependence 
on center of mass energy which means there are less soft 
collision processes as we move up in energy.
  The power $n$ determines if the particle is coming from point like qq scatterings
(leading twist) or from multiple scattering centers involving many 
quarks (higher twists).
 The power $n$ decreases with decreasing $\sqrt{s}$ for all particles in all our 
analysis which indicates that the production process 
is moving from higher twist to leading twist as one goes up in $\sqrt{s}$. 
  The maximum values of $n$ are found to be 16 for proton spectra 12 for 
pion spectra at the lowest RHIC energy. 
  Another important observation is with increasing $\sqrt{s}$, separation 
of powers for protons and for pions narrows down hinting that the baryons and mesons are 
governed by same production process as one moves to the highest LHC energy.
 Separation of $n$ for protons and pions is more at lower energy. This would imply that
higher twist process prefer pion production over proton at lower energies.

\section{Acknowledgements}
  We acknowledge the financial support from Board of Research in Nuclear 
Sciences (BRNS) for this project. Prashant Shukla thanks his CMS colleagues 
for their comments.

\end{document}